\documentclass{aa}  

\usepackage{graphicx}
\usepackage{subfigure}
\usepackage{subfloat}
\usepackage{float}
\usepackage[version=3]{mhchem}
\usepackage{booktabs}
\usepackage{pdflscape}
\usepackage{supertabular}
\usepackage{longtable,lscape}
\usepackage{txfonts}
\begin{document} 

\title{Lupus disks with faint CO isotopologues: \\low gas/dust or large carbon depletion?}


\author{A. Miotello\inst{\ref{inst:leiden}} \and
            E. F. van Dishoeck\inst{\ref{inst:leiden},\ref{inst:mpe}}\and
            J. P. Williams\inst{\ref{inst:Hawaii}}\and
            M. Ansdell\inst{\ref{inst:Hawaii}} \and
            G. Guidi\inst{\ref{inst:INAF}} \and
            M. Hogerheijde\inst{\ref{inst:leiden}} \and
            C. F. Manara\inst{\ref{inst:ESA}} \and \\
            M. Tazzari\inst{\ref{inst:ESO},\ref{inst:EC}} \and
            L. Testi\inst{\ref{inst:INAF},\ref{inst:ESO},\ref{inst:EC}} \and
            N. van der Marel\inst{\ref{inst:Hawaii}} \and 
            S. van Terwisga\inst{\ref{inst:leiden}}}

\institute{
Leiden Observatory, Leiden University, Niels Bohrweg 2, NL-2333 CA Leiden, The Netherlands\label{inst:leiden}\and
Max-Planck-institute f{\"u}r extraterrestrische Physik, Giessenbachstra{\ss}e, D-85748 Garching, Germany\label{inst:mpe}\and
Institute for Astronomy, University of Hawaii at Manoa, 2680 Woodlawn dr., Honolulu, HI, 96822, USA\label{inst:Hawaii}\and
INAF/Osservatorio Astrofisico di Arcetri, Largo E. Fermi 5, I-50125 Firenze, Italy\label{inst:INAF}\and
Scientific Support Office, Directorate of Science, European Space Research and Technology Centre (ESA/ESTEC), Keplerlaan 1, 2201AZ Noordwijk, The Netherlands\label{inst:ESA}\and
European Southern Observatory, Karl-Schwarzschild-Str. 2, D-85748 Garching bei M\"{u}nchen, Germany\label{inst:ESO}\and
Excellence Cluster “Universe”, Boltzmann str. 2, D-85748 Garching bei M\"{u}nchen, Germany\label{inst:EC}}
                       

\abstract{An era has started in which gas and dust can be observed independently in protoplanetary disks, thanks to the recent surveys with the Atacama Large Millimeter/sub-millimeter Array (ALMA). The first near-complete high-resolution disk survey in both dust and gas in a single star-forming region has been carried out in Lupus, finding surprisingly low gas-to-dust ratios \citep[][]{Ansdell16}\textbf{.}} 
{The goal of this work is to fully exploit CO isotopologues observations in Lupus, comparing them with physical-chemical model results, in order to obtain gas masses for a large number of disks and compare gas and dust properties.} 
{We have employed the grid of physical-chemical models presented in \cite{Miotello16} to analyze continuum and CO isotopologues ($^{13}$CO $J=3-2$ and C$^{18}$O $J=3-2$) observations of Lupus disks, including isotope-selective processes and freeze-out. Employing also the ALMA $^{13}$CO-only detections, disk gas masses have been calculated for a total of 34 sources, expanding the sample of 10 disks studied by \cite{Ansdell16}, where also C$^{18}$O was detected.} 
{We confirm that overall gas-masses are very low, often smaller than 1$M_{\rm J}$, if volatile carbon is not depleted. Accordingly, global gas-to-dust ratios are much lower than the expected ISM-value of 100, being predominantly between 1 and 10.  Low CO-based gas masses and gas-to-dust ratios may indicate rapid loss of
gas, or alternatively chemical evolution, e.g. via sequestering of carbon from CO to more complex molecules, or carbon locked up in larger bodies.} 
{Current ALMA observations of $^{13}$CO and continuum emission cannot distinguish between these two hypotheses. We have simulated both scenarios, but chemical model results do not allow us to rule out one of the two, pointing to the need to calibrate CO-based masses with other tracers. Assuming that all Lupus disks have evolved mainly due to viscous processes over the past few Myr, the observed correlation between the current mass accretion rate and dust mass found by \cite{Manara16} implies a constant gas-to-dust ratio, which is close to 100 based on the observed $M_{\rm disk}/\dot{M}_{\rm acc}$ ratio. This in turn points to a scenario in which carbon depletion is responsible for the low CO isotopologue line luminosities.} 


\maketitle
%
\section{Introduction}
Protoplanetary disks have been extensively studied in the past decades with infrared (IR) surveys \citep[][]{Haisch01,Hernandez07,Evans09}.
Furthermore, several attempts have been carried out to study the structure and bulk content of disks independently in gas and dust \citep[see e.g.,][as a review.]{Thi01,Panic08,Panic09,Andrews12,Boneberg16,Andrews15}. However, these studies have been limited to small, non statistically significant samples of mostly Herbig disks, often with partially unresolved and limited sensitivity observations. The Atacama Large Millimeter/sub-millimeter Array (ALMA) now has the sensitivity and resolving power needed to image large numbers of disks in continuum and molecular lines in a modest amount of time. The first near-complete survey of protoplanetary disks in both dust and gas with ALMA in a single star-forming region has been carried out in Lupus \citep[][]{Ansdell16}. For the first time around 80 disks in the same region have been observed at a resolution of 0.3" (22 -- 30 au radius for a distance of 150 -- 200 pc). At the same time the unprecedented sensitivity of ALMA allowed detection of a fraction of these disks in the faint CO isotopologue lines with just one minute observations per source. Moreover, stellar properties and mass accretion rates have been estimated from VLT/X-Shooter spectra for the same sample, allowing to build a complete picture for the Lupus sources \citep[][Alcala et al. subm.]{Alcala14,Manara16}. More recently, other star forming regions have been surveyed with similar aims by ALMA, such as Chameleon I \citep{Pascucci16} and the more evolved Upper Scorpius region \citep{Barenfeld16}. An era has started in which gas and dust can be observed independently in protoplanetary disks.

Together with the gas physical structure, the total gas mass is one of the crucial properties needed for describing disk evolution. Starting from the process of grain-growth, planetesimal formation sensitively depends on the physical structure of the gaseous disk \citep[see e.g.][for a review]{Armitage11}. For this reason it is crucial to directly observe the bulk of the gas in protoplanetary disks. CO lines are commonly used to estimate the total disk gas mass.  In particular, less abundant CO isotopologues, which become optically thick deeper in the disk than $^{12}$CO, can trace the bulk gas mass present in the molecular layer \citep{vanZadelhoff01,Dartois03}. 

The main caveat when employing CO isotopologues as mass tracers is related to the conversion of the observed CO mass into total gas mass. There are various arguments why CO/H$_2$ is less than the canonical value of $10^{-4}$ \citep{Aikawa97} when averaged over the entire disk as indicated by early observatios \citep[][]{Dutrey97}. Processes like CO photodissociation in the upper layers and freeze-out in the disk's midplane have been identified as the main cause \citep{vanZadelhoff01,Aikawa02} and these processes are well understood from a molecular physics point of view. They are included in all recent physical-chemical disk models \citep[][]{Hollembach05,Gorti09,Nomura09,Woitke09,Bruderer12,Bruderer13}. When less abundant CO isotopologues are involved, the well understood process of isotope-selective photodissociation, destroying relatively more C$^{18}$O, needs to be taken into account \citep{Visser09,Miotello14,Miotello16}.

\cite{Williams14} have shown that by combining multiple CO isotopologues, such as  $^{13}$CO and C$^{18}$O,  and accounting in a parametrized way for photodissociation and freeze-out, it is possible to estimate disk gas masses, irrespective of the disk properties. Traditionally, then a constant ISM-like overall volatile carbon abundance is used to calculate the total H$_2$ mass ($X_{\rm C}=$[C]/[H]=1.35$\times 10^{-4}$). This assumption may however be incorrect, as high levels of carbon depletion have been inferred for at least one disk. This is the case for TW Hya, the closest and probably best studied disk, for which the fundamental rotational transition of hydrogen deuteride (HD) has been observed by the \emph{Herschel Space Observatory} \citep{Bergin13}. Comparing with SMA C$^{18}$O data, \cite{Favre13} found that a two orders of magnitude carbon depletion was needed in order to recover the HD-based disk mass determination from C$^{18}$O. This result has been confirmed by physical-chemical modeling of the source, which was able to reproduce, among other lines, spatially resolved ALMA data and atomic carbon lines \citep{Kama16,Schwarz16}. This result is interpreted as chemical evolution, i.e., carbon has been turned from CO into more complex species either in the gas or in the ice \citep{Aikawa96,Bergin14,Drozdovskaya15,Eistrup16} or as a grain growth effect, carbon has been locked up in large icy bodies that no longer participate in the gas-phase chemistry \citep{Du15,Kama16} resulting in less bright CO lines. This hints to the possible existence of a class of disks where CO is not the main carbon reservoir. 

In this paper we apply the detailed modeling technique developed by \cite{Miotello16} to the ALMA $^{13}$CO and C$^{18}$O Lupus observations. Our modeling procedure allows us to employ also the $^{13}$CO-only detections to provide disk gas mass determinations, for a total of 34 disks, extending and refining the initial analysis presented in \cite{Ansdell16} for only 10 disks, for which both $^{13}$CO and C$^{18}$O lines are available. Our derived gas masses are generally very low, often lower than the mass of Jupiter, consistent with \cite{Ansdell16}. This translates to very low global gas-to-dust ratios, often approaching unity (Sec. \ref{discussion}). Low CO-based gas masses and gas-to-dust ratios may indicate rapid disk evolution, which is usually taken to be a loss of gas. An alternative is that the CO abundance may be low, as discussed above. Since current data cannot distinguish between these scenarios, we have simulated both of them, but model results do not allow us to rule out one of the two. Future ALMA observations of more complex tracers are needed to calibrate CO-based masses. Alternatively, \cite{Manara16} and \cite{Rosotti16} have proposed a method that allows to trace the gaseous disk component, independently from chemistry by exploiting the availability of dust disk mass and mass accretion rates measurement. The implications of these data are discussed in Sec. \ref{Macc}. 

\section{ALMA observations}

Lupus is one of the youngest \citep[$3 \pm 2$ Myr;][]{Alcala14} and closest complex, composed of four main star-forming regions \citep[Lupus I-IV, see][for a review]{Comeron08}.  Lupus III is located at a distance of $\sim 200$ pc to the Sun, while Lupus I, II, and IV at $\sim 150$ pc. 

The sample employed for this paper was observed in the Cycle 2 \emph{Lupus ALMA  Disk Survey}, it is composed of 88 objects with $0.1 < M_{\star}/M_{\odot} < 2.84$ and it is presented in detail by\cite{Ansdell16}.
The observations were obtained on 2015 June 14 and 15. The disks were observed in continuum (890 $\mu$m) and CO isotopologues line emission ($^{13}$CO and C$^{18}$O 3-2 transitions). More details on the observational settings and data reduction are presented in \cite{Ansdell16}.
Of the 88 targets, 34 were detected in $^{13}$CO while only 10 were detected in C$^{18}$O at more than a 3$\sigma$ level.

\section{Model}

The observed CO isotopologue fluxes have been compared with the model results presented in \cite{Miotello16}.
These models are computed with the physical-chemical code DALI, extensively tested 
with benchmark test problems \citep[][]{Bruderer12,Bruderer13} and against 
observations \citep[][]{Bruderer12,Fedele13,Bruderer14}.  
The dust temperature, $T_{\rm dust}$, and local continuum radiation field from UV to mm wavelengths are calculated by a 2D Monte Carlo method, given an input disk density structure and a stellar spectrum. Then the chemical composition is obtained with a time-dependent chemical network simulation (1 Myr). 
Finally, the gas temperature, $T_{\rm gas}$, is obtained from an iterative balance between heating and cooling processes until a self-consistent solution is found and the non-LTE excitation of the molecules is computed. The final outputs are spectral image cubes created with a raytracer. As in \cite{Miotello14} and in \cite{Miotello16}, a complete treatment of CO isotope-selective 
processes is included. 

The grid by \cite{Miotello16} is composed of around 800 disk models that have been run for a range of realistic disk and stellar parameters. Disk masses vary between $10^{-5} M_{\odot}$ and $10^{-1} M_{\odot}$ and stellar luminosities are set to 1 $L_{\odot}$ (plus excess UV for mass accretion rates of $10^{-8} M_{\odot} \,\rm yr^{-1}$) for T Tauri disks and 10 $L_{\odot}$ for Herbig disks \citep[see Table 1 in][]{Miotello16}. Coupling of the UV excess to an accretion rate is merely a convenient prescription to obtain a value for the UV luminosity, which controls the thermal and chemical structure of the gas including the number of CO dissociating photons (Kama et al. 2016). It should not be taken here as a measure for disk evolution timescales. A consistent description of the accretion process and disk evolution is out of the scope of our models.
Line luminosities for different CO istopologues and transitions have been ray traced. The majority of Lupus sources detected in both CO isotopologues have stellar luminosity $L_{\star}$ between 0.1 $L_{\odot}$ and 1 $L_{\odot}$ (Alcala et al., subm.). As a check, a smaller grid of T Tauri disk models with $L_{\star}=0.1 \,L_{\odot}$ and excess UV for mass accretion rates of $10^{-8} M_{\odot}$ yr$^{-1}$, that generates about 0.2 $L_{\odot}$ of UV accretion luminosity, has been run to reproduce the extreme cases. However, we found that CO isotopologues line luminosities are reduced only up to 25\%. As this is well within the uncertainties on the CO-based gas mass derivations by \cite{Miotello16}, the original grid of T Tauri disk models with $L_{\star}=1 \,L_{\odot}$ has been used for the analysis.

\section{Results}
\label{Results}

\subsection{Dust masses revisited}
\label{dust_masses}
Continuum observations at sub-mm wavelengths are traditionally employed to trace large (mm-sized) grains present in the cold midplane of disks. Since the bulk of the dust mass is expected to be retained in large grains, their thermal emission is generally used to derive disk dust masses, following:
\begin{equation}
M_{\rm dust}=\frac{F_{\nu} \, d^2}{\kappa_{\nu}B_{\nu}(T_{\rm dust})}.
\label{eq_Mdust}
\end{equation}
The sub-mm flux $F_{\nu}$ is directly related to the dust mass $M_{\rm dust}$ under the assumption that the emission is optically thin and in the Rayleigh-Jeans regime \citep{Beckwith90}. Eq. \ref{eq_Mdust} has been employed by \cite{Ansdell16} to derive dust masses for the Lupus disks, assuming a dust opacity $\kappa_{\nu}=3.4$ cm$^2$g$^{-1}$ at 340 GHz and a characteristic dust temperature $T_{\rm dust}=20$ K. This first-order analysis does not account for the effects of disk inclination and temperature variation on the spatially integrated dust emission. However, the inclination angle $i$ has been derived for many of the Lupus disks from continuum data by \cite{Ansdell16} and Tazzari et al (in prep.).

In this work we exploit the results obtained by \cite{Miotello16} with DALI continuum radiative transfer to estimate Lupus disks dust masses, accounting for the disk inclination and temperature structure. The assumed dust opacity,  $\kappa_{\nu}=4.3$ cm$^2$g$^{-1}$ at 340 GHz, is the standard value in DALI \citep[][]{WD01,Bruderer13} and is slightly higher than that used by \cite{Ansdell16}. Our models assume that the dust surface density distribution follows a radial power-law with an exponential taper and a Gaussian distribution in the vertical direction. Two dust populations are considered: large (mm-sized) grains are settled toward the midplane and small (sub-$\mu$m to $\mu$m-sized) grains are coupled to the gas and present in the disk atmosphere \citep[see][for more detail]{Miotello16}. Finally, grain-growth and migration are not included in the models, except in an ad-hoc fashion by including small and large grains.  

Both the continuum and line emission obtained with the disk models presented in \cite{Miotello16} have been ray-traced assuming disk inclinations of 10$^{\circ}$ and 80$^{\circ}$. Sub-millimeter fluxes are not expected to vary significantly for intermediate inclinations, i.e., up to 70$^{\circ}$ \citep{Beckwith90}. On the other hand, line intensities and dust emission are extremely modified in the rarer case of edge-on disks, i.e., with an inclination close to 90$^{\circ}$. Lupus disks with $i<70^{\circ}$ or with unknown inclination are compared with models where $i$ = 10$^{\circ}$, elsewhere we employ models with $i$ = 80$^{\circ}$.

\begin{figure}
   \resizebox{\hsize}{!}
             {\includegraphics[width=1.\textwidth]{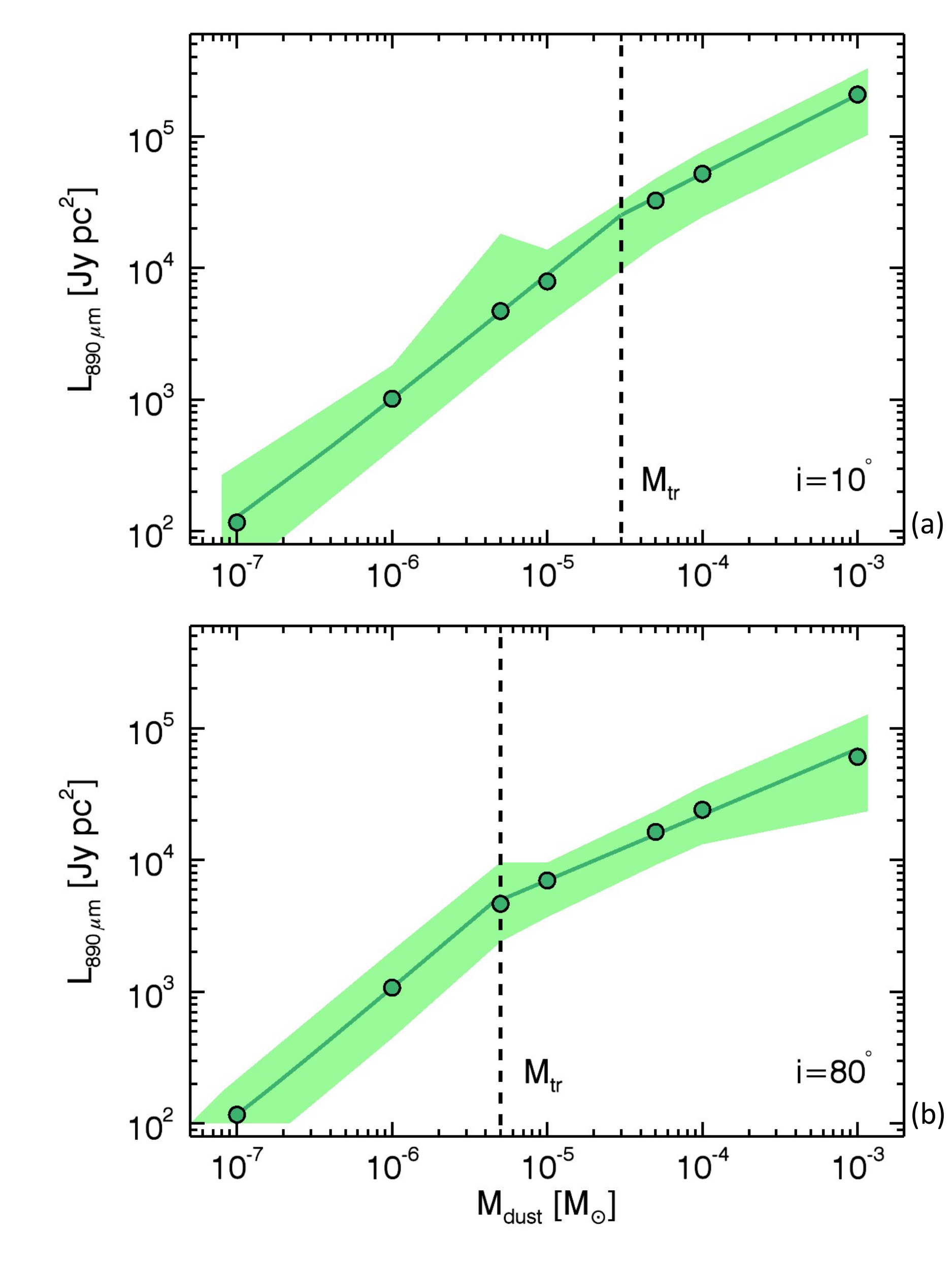}}
      \caption{Medians of the simulated T Tauri continuum luminosities at 890 $\mu$m for different mass bins, presented by the green dots for disk inclinations of 10$^{\circ}$ and 80$^{\circ}$ in panel (a) and (b) respectively. The solid lines show the fit function of the 890 $\mu$m continuum luminosity as a function of the disk mass (see Eq. \ref{Lcont_Mdust}). The dotted black lines present the transition mass between the two dependencies of the luminosity on the mass expressed in Eq. \ref{Lcont_Mdust}. The colored bands show the maximum and minimum simulated continuum luminosities for the different mass bins.}
       \label{dust_fit}
\end{figure}

The medians of the simulated continuum luminosities at 890 $\mu$m can be calculated for different dust masses (Fig. \ref{dust_fit}). They can be fitted by simple functions of the disk dust mass and be employed to estimate Lupus disk masses. In particular, the continuum luminosity at 890 $\mu$m $L_{\rm 0.9mm}$ can be expressed by:
\begin{equation}
L_{\rm 0.9mm}=
  \begin{cases}
a+b\cdot M_{\rm dust}^c & \quad M_{\rm dust} < M_{\rm tr}\\
d + e\cdot M_{\rm dust}^f & \quad M_{\rm dust} > M_{\rm tr}.\\
  \end{cases}
\label{Lcont_Mdust}
\end{equation}

The coefficients $a, b, c, d, e, f,$ and $M_{\rm tr}$ are reported in Table \ref{coeff_cont} for disk inclinations of both 10$^{\circ}$ and 80$^{\circ}$. For dust masses below the transition mass $M_{\rm tr}$ the dependence is linear, while for higher masses it is sub linear. This reflects the fact that the dust emission becomes optically thick for high enough dust masses, even at 890 $\mu$m. This occurs at lower masses for edge-on disks, as the high inclination enhances the dust column density.

\begin{table}[tbh]
\caption{Polynomial coefficients $a, b, c, d, e, f,$ and $M_{\rm tr}$, in Eq. (\ref{Lcont_Mdust})}
\label{coeff_cont}
\centering
\begin{tabular}{ccc}
\toprule
&$i=10^{\circ}$ &$i=80^{\circ}$\\
\midrule
$a$&$17$&$10$\\
$b$&$5 \cdot 10^{8}$&$10^{9}$\\
$c$&$0.95$&$1.0$\\
$d$&$76$&$40$\\
$e$&$1.3 \cdot 10^{7}$&$2.2 \cdot 10^{6}$\\
$f$&$0.6$&$0.5$\\
$M_{\rm tr} [M_{\odot}]$&$3 \cdot 10^{-5}$&$5 \cdot 10^{-6}$\\
\hline
\end{tabular}
\end{table}

\begin{figure}
   \resizebox{\hsize}{!}
             {\includegraphics[width=1.\textwidth]{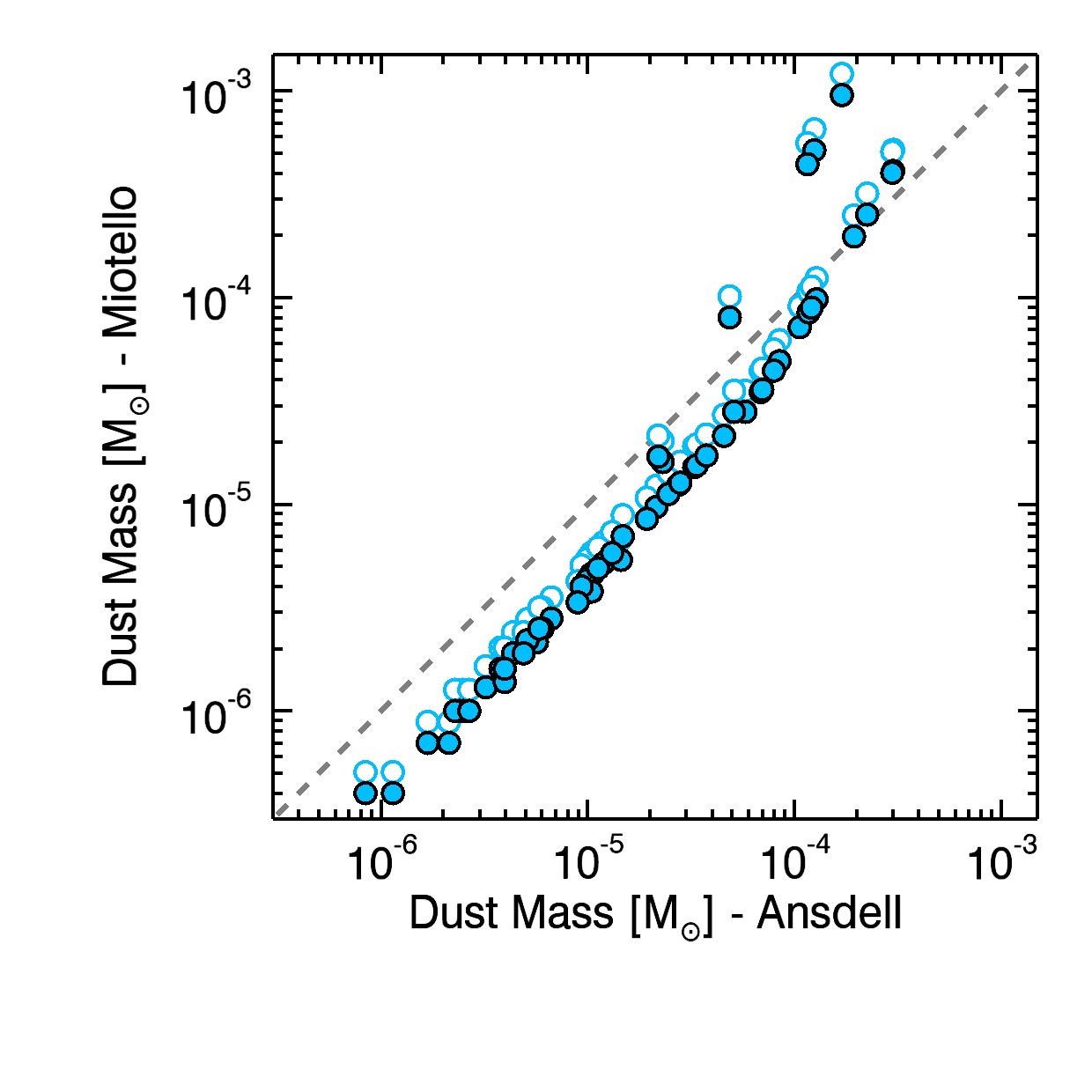}}
      \caption{Disk dust masses derived in this work compared with the dust masses obtained by \cite{Ansdell16}. Full symbols show the current model results. Empty circles present mass estimates corrected for the dust opacity difference at 340 GHz between the two studies.}
       \label{compare_megan}
\end{figure}

The Lupus disk dust masses derived in this work are generally a factor of a few lower than the estimates given by \cite{Ansdell16} (see Fig. \ref{compare_megan}). In particular, between $5\cdot 10^{-7} M_{\odot}$ and $3\cdot 10^{-5} M_{\odot}$, dust masses derived by \cite{Ansdell16} are a factor of 2.2 higher than those estimated by this analysis (1.7 if we correct for the dust opacity difference at 340 GHz). For disks with dust masses larger than $M_{\rm tr}=3\times10^{-5} M_{\odot}$ the derivations obtained with the two methods become similar because the emission becomes marginally optically thick. This effect is only considered by our models and leads to higher mass determinations. Finally a bi-modality can be seen in Fig. \ref{compare_megan}, with six points being displaced from the main trend. These represent the highly inclined disks ($i>70^{\circ}$) with dust masses larger than $M_{\rm tr}=5\times10^{-6} M_{\odot}$. For the most massive of these sources, the dust masses derived in this work are generally a factor of a few higher than the estimates given by \cite{Ansdell16}. The dust masses derived for the highly inclined disks are $M_{\rm dust}= 9.6\cdot10^{-4},\, 5.2\cdot10^{-4},\, 4.4\cdot10^{-4},\, 8.0\cdot10^{-5},\, 1.7\cdot10^{-5},\, 1.6\cdot10^{-5},\, 7.0\cdot10^{-6},\,  3.4\cdot10^{-6},\, 2.2\cdot10^{-6}
M_{\odot}$ for J16083070-3828268, MY Lup, J16070854-3914075, Sz 133, J16090141-3925119, Sz 84, Sz 74, J16102955-3922144, and J16070384-3911113 respectively.

\subsection{Gas masses}
\label{gas_masses}
\begin{figure}
   \resizebox{\hsize}{!}
             {\includegraphics[width=1.\textwidth]{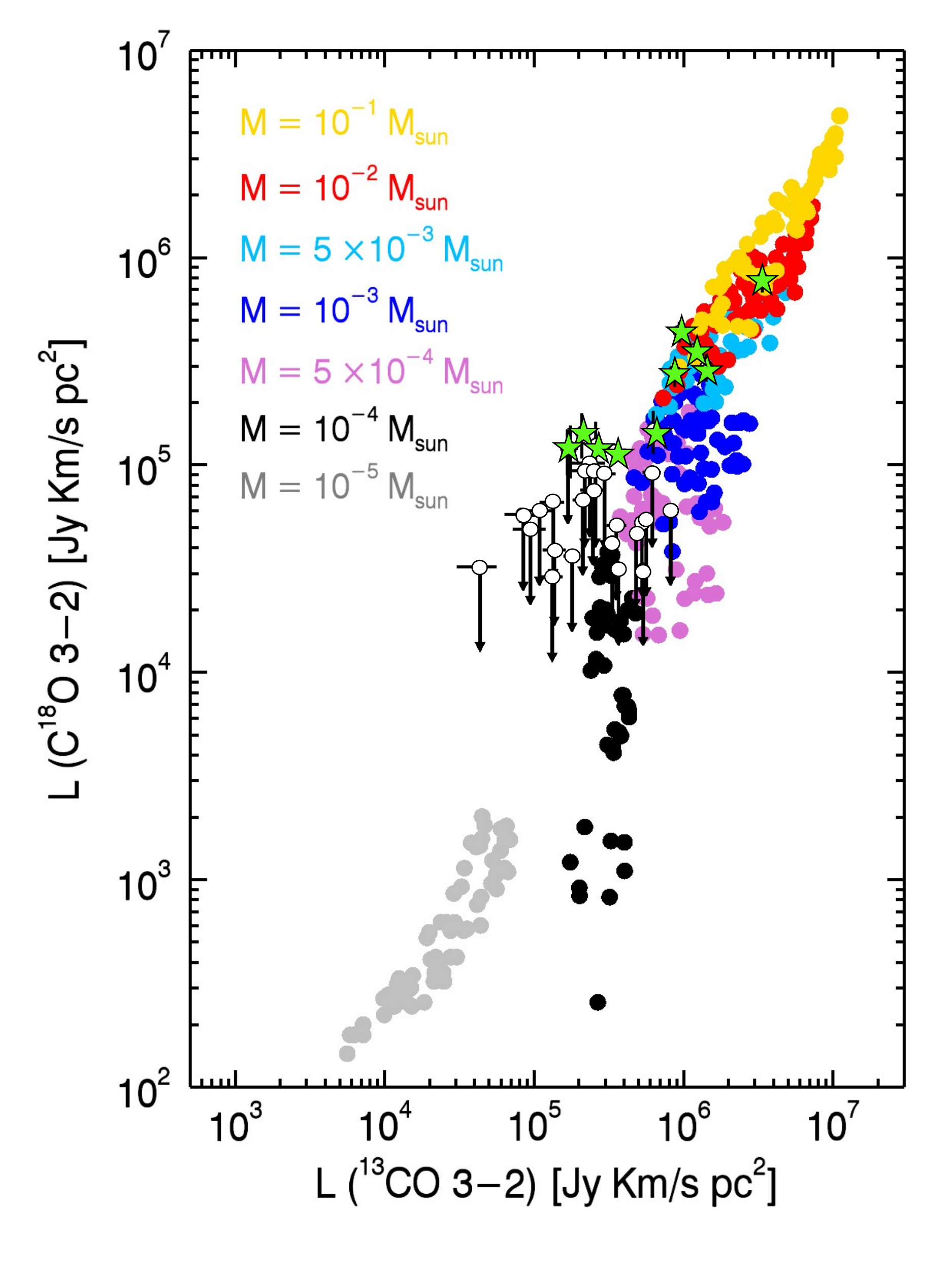}}
      \caption{C$^{18}$O (3-2) vs $^{13}$CO (3-2) line luminosities. The color-coded dots show the line luminosities simulated by the T Tauri models in the \cite{Miotello16} grid. The Lupus observations are overplotted. The disks detected in both isotopologues are shown by the green stars, while the C$^{18}$O non detections are presented by the empty circles.}
       \label{largegrid}
\end{figure}

Measuring disk gas masses is essential for understanding disk evolution up to the formation of planetary systems. The aim of this analysis is to employ the CO isotopologues lines observed in Lupus for deriving gas masses for a statistically significant number of disks.
The $^{13}$CO and C$^{18}$O model results shown by \cite{Miotello16} can be presented in the same way as done by \cite{Williams14} and be compared with Lupus observations (Fig. \ref{largegrid}). As explained by \cite{Miotello16} the two sets of models results differ over the whole disk mass range because of a temperature effect. 
\cite{Williams14} find a wider range of CO luminosities because they parameterize a wider range of temperatures than what \cite{Miotello16} find in their models which compute the disk temperature structure through full radiative transfer.
Furthermore, the divergence is maximized in the lower mass regime where the implementation of isotope-selective processes reduces C$^{18}$O line intensities compared with the results by \cite{Williams14}.
Same disk mass model results cover a limited region of the line luminosity-luminosity space. Lupus disks that are detected in both isotopologues are presented in Fig. \ref{largegrid} with the green stars, while C$^{18}$O non-detections are shown with the white circles as upper limits on the y direction. Such a plot can be used to derive disk masses only if also C$^{18}$O is detected. In the case of Lupus this is the case for only 10 sources. Four of them are not reproduced by any of our models, presenting either lower $^{13}$CO and regular C$^{18}$O luminosities, or regular $^{13}$CO and higher C$^{18}$O luminosities (see Fig. \ref{largegrid}). This shows that the current model grid may not apply to all disks, but a source-by-source detailed modeling may be needed. Knowledge about disk properties such as the radial extent and vertical structure would help in a more secure prediction of the line fluxes. 

\begin{figure}
   \resizebox{\hsize}{!}
             {\includegraphics[width=1.\textwidth]{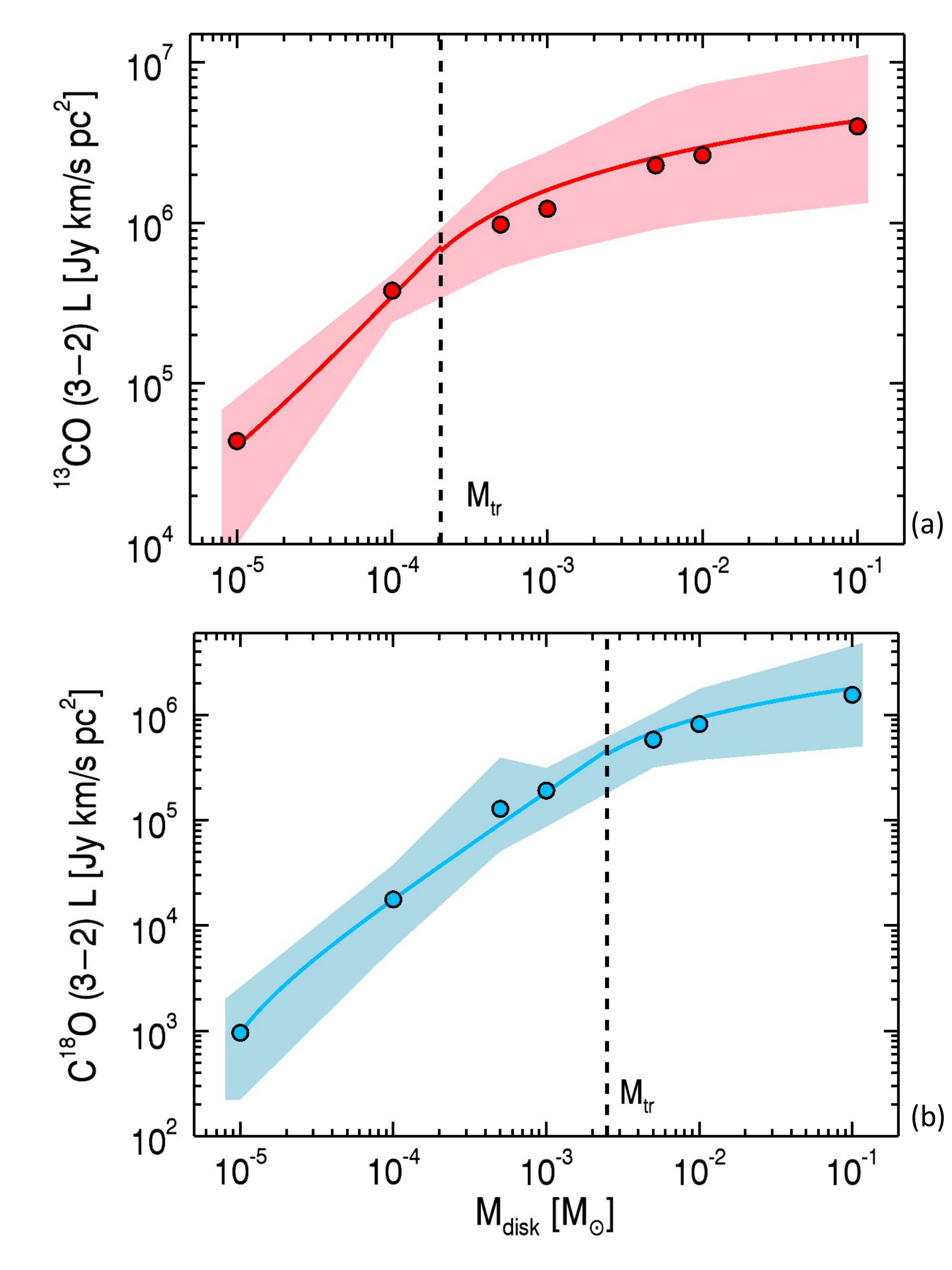}}
      \caption{Medians of the $^{13}$CO (3-2) and C$^{18}$O (3-2) simulated T Tauri line luminosities for different mass bins, presented by the red and blue dots in panel (a) and (b) respectively ($i=10^{\circ}$). The red and blue lines show the fit function of the line luminosity as function of disk mass. The dotted black lines present the transition mass between the linear and the logaritmic dependence of the line uminosity on the mass (see Eq. \ref{L_13}). The colored bands show the maximum and minimum simulated line luminosities for the different mass bins.}
       \label{fit}
\end{figure}
\begin{figure}
   \resizebox{\hsize}{!}
             {\includegraphics[width=1.\textwidth]{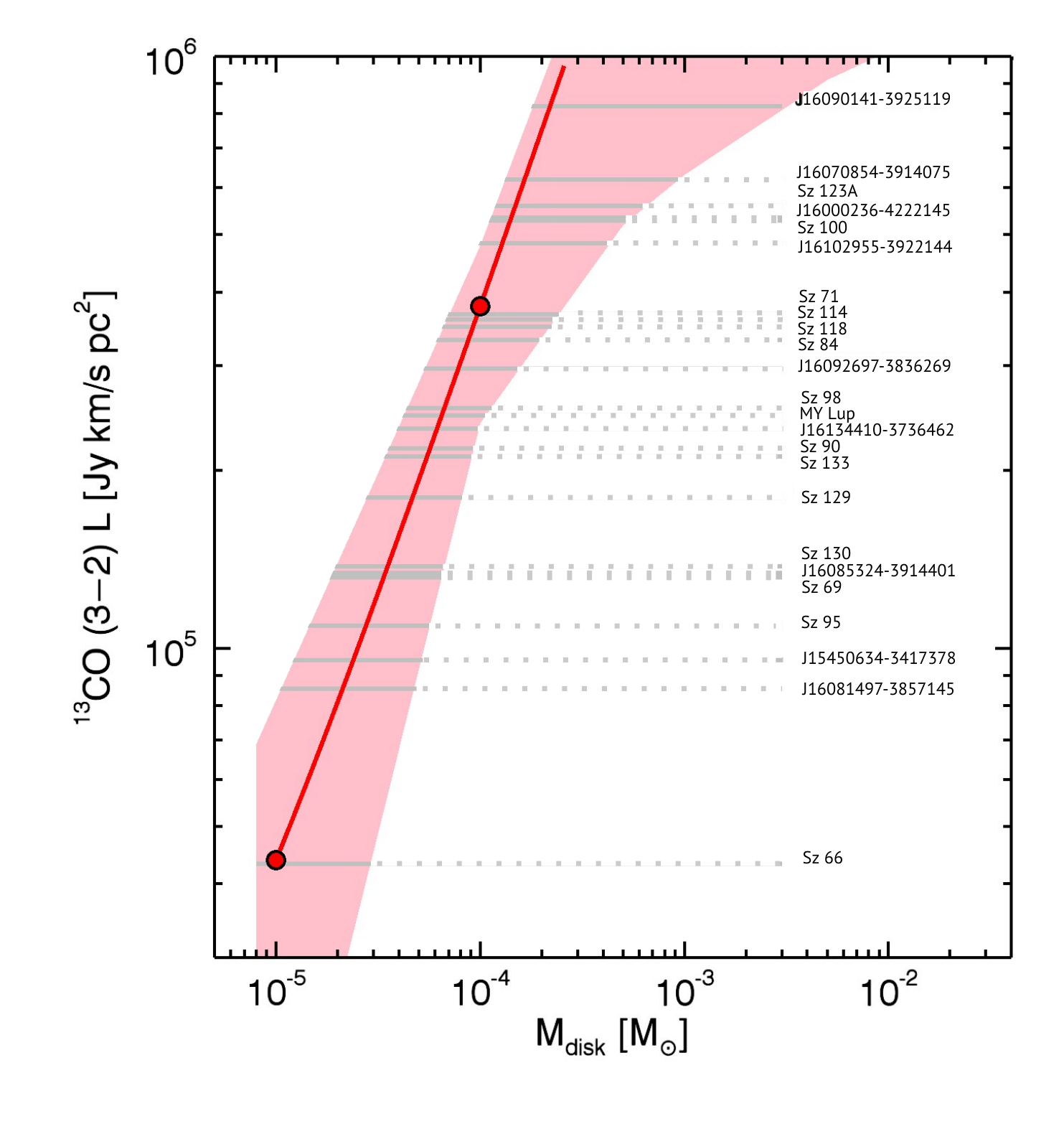}}
      \caption{Blow-up of Fig. 4 top: medians of the $^{13}$CO (3-2) simulated line luminosities for different mass bins, presented by the red dots in the region where the luminosity dependence on mass is linear (see Eq. \ref{L_13}). The red line shows the fit function of the line luminosity as function of disk mass. The $^{13}$CO-only detections are shown by the gray dashed lines. The pink band shows the maximum and minimum simulated line luminosities.}
       \label{13CO_zoom}
\end{figure}

For disks that have been detected in $^{13}$CO, but not in C$^{18}$O (henceforth, $^{13}$CO-only detections), $^{13}$CO alone can be employed as a mass tracer only if its emission is optically thin. Similarly to \cite{Miotello16} the median of the $^{13}$CO and C$^{18}$O $J=3-2$ simulated line luminosities, rather than $J=2-1$, have been expressed by fit functions of the disk mass:
\begin{equation}
L_{y} =
  \begin{cases}
A_y+B_y\cdot M_{\rm gas} & \quad M_{\rm gas} \leq M_{\rm tr}\\
C_y + D_y\cdot  \text{log}_{10}(M_{\rm gas}) & \quad M_{\rm gas} > M_{\rm tr},\\
  \end{cases}
\label{L_13}
\end{equation}
where $y=13$ or 18, for $^{13}$CO and C$^{18}$O respectively. For low mass disks the line luminosity has a linear dependence on the disk mass, while for more massive disks the trend is logarithmic, due to optical depth. The transition point $M_{\rm tr}$ is different for the two isotopologues because $^{13}$CO becomes optically thick at lower column densities than C$^{18}$O. 
The polynomial coefficients $A_y$, $B_y$, $C_y$, and $D_y$, as well as the transition masses $M_{\rm tr}$ are reported in Table \ref{A_mass}, both for disk models with inclination angle $i=10^{\circ}$ and $i=80^{\circ}$. 
The fit functions are shown in Fig. \ref{fit} by the red and blue lines for $^{13}$CO and C$^{18}$O respectively. 

Interestingly, all the $^{13}$CO-only detections fall in the region of panel (a) where the dependence is linear, i.e. for $M_{\rm disk}<M_{\rm tr}$ (Fig. \ref{g_t_d}). Accordingly we can use the fit function presented in Eq. \ref{L_13} to calculate the gas masses of these sources, as $^{13}$CO is optically thin for the observed range of luminosities, confirmed by the absence of C$^{18}$O emission. The uncertainties on the gas mass determinations are defined by the shadowed region in Fig. \ref{13CO_zoom} (that is a blow-up of Fig. \ref{fit}), which covers the range of line luminosities simulated by the grid of models for different mass bins. 
Similarly we can employ this line luminosity-mass relation to derive mass upper limits for disks that are undetected in both CO isotopologues. The disk mass determinations together with the calculated upper limits are presented in Table \ref{tab_masses} and shown in the middle panel of Fig. \ref{g_t_d}. Total gas masses are generally low, often smaller than 1 $M_{\rm Jup}$. The implications of this result are discussed in Sec. \ref{discussion}.

\begin{table}[tbh]
\caption{Polynomial coefficients $A_{y}$, $B_{y}$, $C_{y}$, and $D_{y}$, in Eq. (\ref{L_13})}
\label{A_mass}
\centering
\begin{tabular}{lcc}
\toprule
&$i=10^{\circ}$&$i=80^{\circ}$\\
\cmidrule(lr){2-3}
$A_{13}$&$6.707 \cdot 10^{3}$&$-1.182 \cdot 10^{4}$\\
$B_{13}$&$3.716 \cdot 10^{9}$&$3.110 \cdot 10^{9}$\\
$C_{13}$&$6.066 \cdot 10^{6}$&$5.945 \cdot 10^{6}$\\
$D_{13}$&$1.441 \cdot 10^{6}$&$1.460 \cdot 10^{6}$\\
$M_{\rm tr} [M_{\odot}]$&$2 \cdot 10^{-4}$&$2 \cdot 10^{-4}$\\
\cmidrule(lr){2-3}
$A_{18}$&$-9.006 \cdot 10^{2}$&$-7.972 \cdot 10^{2}$\\
$B_{18}$&$51.853 \cdot 10^{8}$&$1.120 \cdot 10^{8}$\\
$C_{18}$&$2.653 \cdot 10^{6}$&$2.581 \cdot 10^{6}$\\
$D_{18}$&$8.500 \cdot 10^{5}$&$8.900 \cdot 10^{5}$\\
$M_{\rm tr} [M_{\odot}]$&$2.5 \cdot 10^{-3}$&$3.0 \cdot 10^{-3}$\\
\hline
\end{tabular}
\end{table}

\section{Discussion}
\label{discussion}

\subsection{Gas-to-dust ratio}
\begin{figure*}
   \resizebox{\hsize}{!}
             {\includegraphics[width=1.\textwidth]{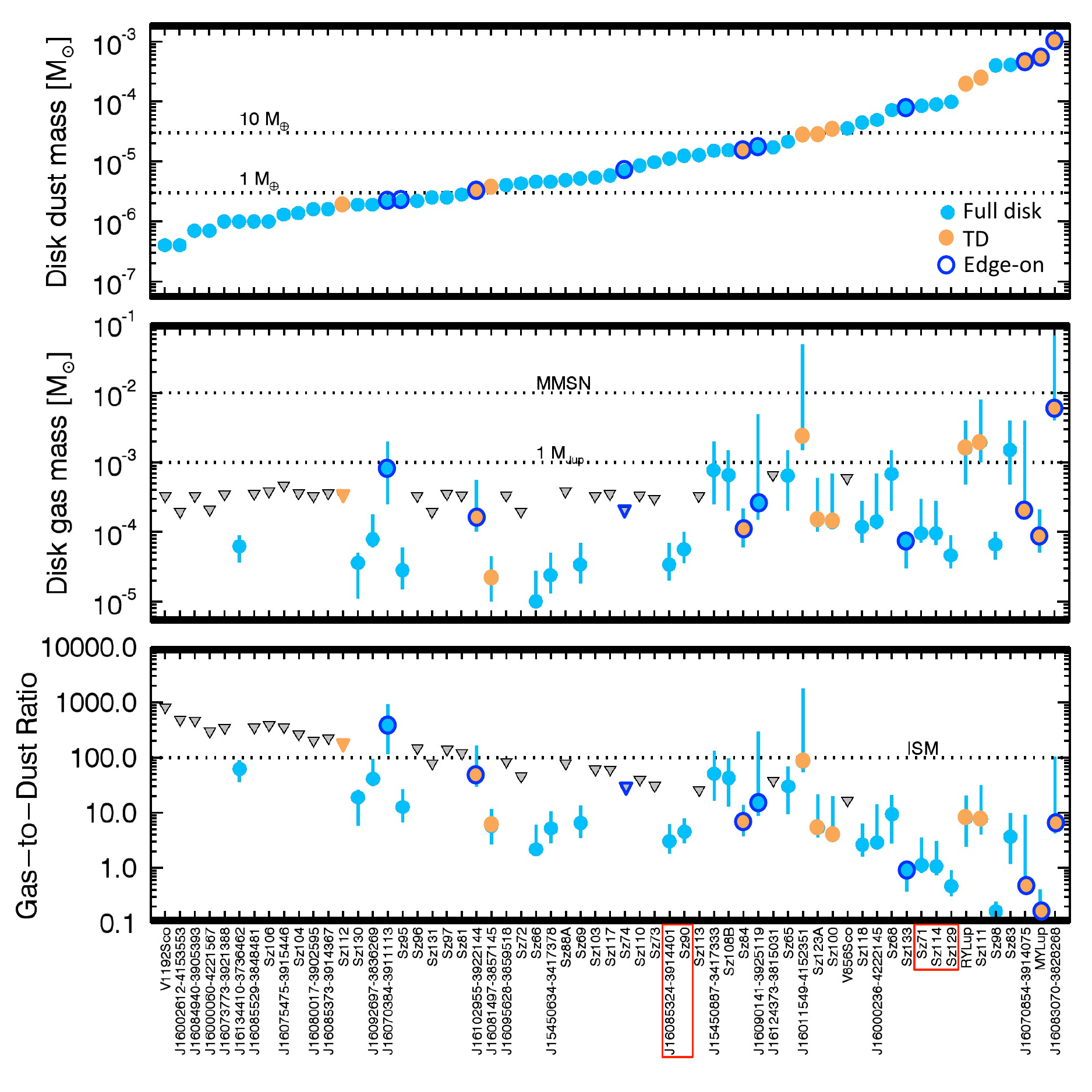}}
      \caption{Disk dust masses (upper panel), gas masses (middle panel), and gas-to-dust ratios (lower panels) for all the sources detected in the continuum. Gas non-detections are shown by gray triangles, while all detections are presented by filled circles. Objects that are classified as transitional disks (TD) are shown in orange and edge on disks are circled in blue. Red rectangles show the subsample of disks discussed in Sec. \ref{test_models}.}
       \label{g_t_d}
\end{figure*}

Dust masses have been calculated for all Lupus disks detected in the continuum as described in Sect. \ref{dust_masses}. For the 34 sources for which at least one of the two CO istopologues was detected, disk gas masses have been derived as explained in Sect. \ref{gas_masses}. Dividing the gas masses by the newly determined dust masses it is possible to obtain global gas-to-dust mass ratios. Fig. \ref{g_t_d} shows that these are often much lower than the expected ISM value of 100, occasionally reaching unity or less. 

Only full disk models have been employed for the mass determinations, but in fact, in the Lupus sample three disks show resolved dust cavities and  three other sources show possible cavities with diameter $\lesssim 0.4"$ \citep{Ansdell16}. Moreover, six other disks are classified as transition disks candidate, but do not show cavities in the ALMA images \citep{Merin10,Romero12,vdMarel16,Bustamante15}. All these sources, presented by orange symbols in Fig. \ref{g_t_d}, are not properly described by our grid of full disk models. However, these moderate resolution Lupus data trace primarily the outer disks, which should be well represented by a full disk model even for transitional disks. Therefore the calculated gas-to-dust ratios provide a first order description of the disk properties in Lupus including the transitional disks. 

\begin{figure}
   \resizebox{\hsize}{!}
             {\includegraphics[width=1.\textwidth]{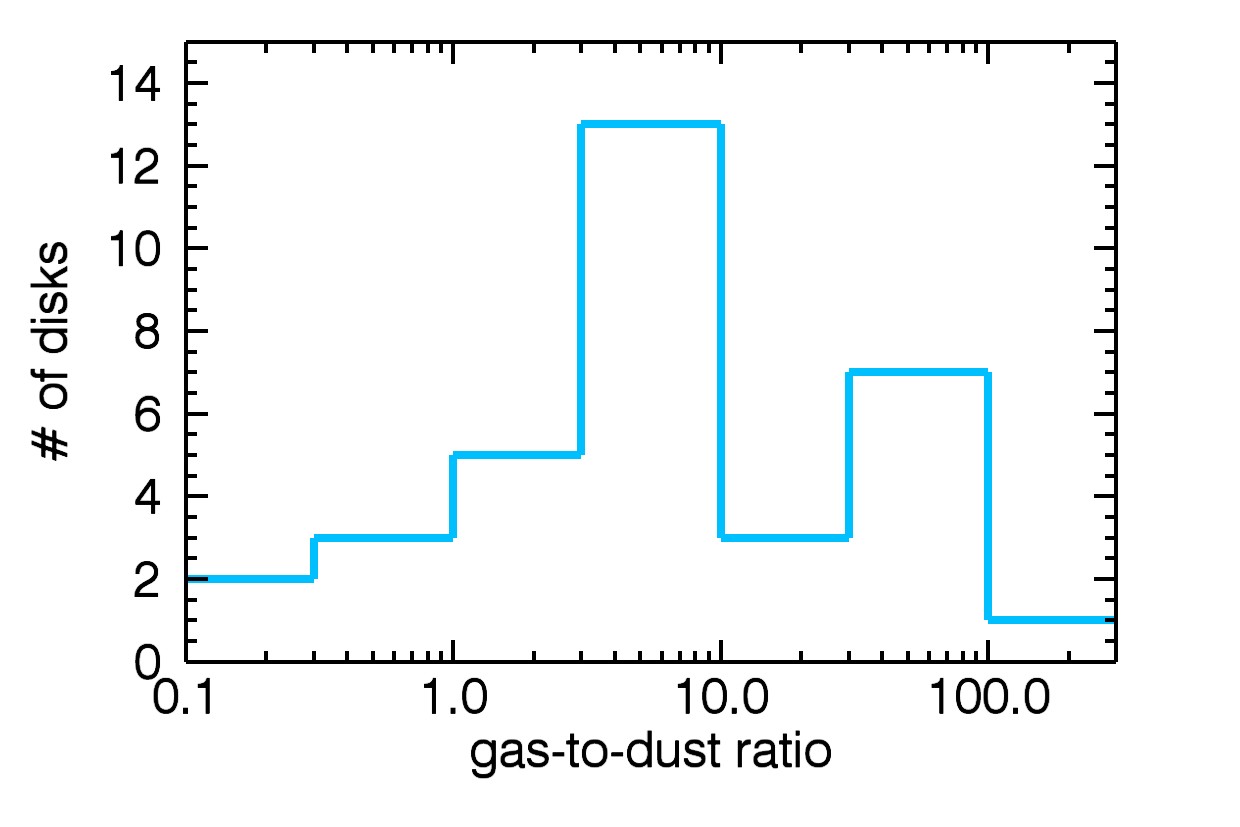}}
      \caption{Histogram showing the number of disks presenting different levels of gas-to-dust ratio. Only sources detected both in continuum and line emission have been considered.}
       \label{histo}
\end{figure}

An alternative way to present the results is shown in Fig. \ref{histo}. The global gas-to-dust ratios obtained for the sources detected in both gas and dust are shown by a histogram. Most of the disks, 23 out of 34, present gas-to-dust ratios lower than 10, with 13 of these sources showing ratios between 3 and 10.

Traditionally disk masses are thought to be dominated by the gaseous component with ISM-like gas-to-dust ratios of 100 assumed to convert $M_{\rm dust}$ into total disk mass. Many disks in Lupus have instead smaller measured gas-to-dust ratios, as shown in the bottom panel of Fig. \ref{g_t_d}. A similar result was found by \cite{Ansdell16} with only 10 gas mass determinations, and it is confirmed with a larger sample of 34 gas mass measurements. A possible interpretation is that Lupus disks are evolved and that the gas has been physically dissipated, while the large dust grains are still retained in the midplane. The finding of disks already depleted in gas by a few Myr would put strong constraints on disk evolution and planet formation theories \citep{Thommes08,Lissauer09,Levison15}, as discussed by \cite{Ansdell16}.  

\subsection{Carbon depletion vs low gas masses}
\label{C_depl}

An alternative interpretation of low CO-based gas masses and gas-to-dust ratios is that CO abundance is low, e.g. via
sequestering of carbon from CO to more complex molecules or being locked-up into larger bodies
\citep{Aikawa96,Bergin14,Du15,Eistrup16,Kama16,Yu16}. However, current data cannot distinguish
between these scenarios. 

Clues on tracers to probe these two cases come from the well studied
TW Hya disk. In this unique case, HD far-infrared emission has been employed to determine independently an
accurate gas mass \citep{Bergin13} and thereby calibrate the weak C$^{18}$O
detection. These observations have been interpreted to imply a much
lower abundance of CO, caused by two orders of magnitude carbon
depletion \citep{Favre13,Schwarz16}. This finding has been confirmed
by an independent analysis, where other lines such as [CI], [CII],
[OI], C$_2$H, and the CO ladder have been self-consistently fitted
\citep[][]{Kama16}. Unfortunately there is no current facility
able to detect the fundamental HD transition in the Lupus
disks. However detection of more complex carbon-bearing species can help to
disentangle between the two cases of gas-poor versus carbon-poor disks.

There is some debate on the mechanism(s) responsible for carbon depletion in protoplanetary disks.
A possible explanation comes from gas-phase reactions initiated by X-ray and cosmic ray
ionization of He in the disk. The resulting He$^+$ atoms can react
with gaseous CO and gradually extract the carbon, which can then be
processed into more complex molecules that can freeze onto cold dust
grains at higher temperatures than CO
\citep[][]{Aikawa97,Bruderer12,Favre13,Bergin14,Kama16,Yu16}. Moreover, oxygen will also be
removed from the gas due to freeze out of H$_2$O, CO$_2$ and CO, even
more than carbon \citep{Hogerheijde11,Oberg11,Walsh15}. Accordingly, a way to test the
level of carbon depletion in disks is to compare observations of CO
isotopologues with species like C$_2$H and c-C$_3$H$_2$, whose
gas-phase abundances are sensitive to the gaseous carbon abundance and
[C]/[O] ratio. Indeed, C$_2$H is observed to have very strong emission
in the TW Hya disk \citep{Kastner2015} and is particularly
strong when both elements are depleted but gaseous [C]/[O]$>$1 \citep{Kama16}.  If the difference in the global gas-to-dust ratios found in Lupus
disks is due to different levels of carbon
depletion, this should be reflected in C$_2$H and c-C$_3$H$_2$ fluxes and this may be tested by future ALMA observations. Alternatively, ice chemistry may be the fundamental process turning CO in more complex organics, such CH$_3$OH, or in CO$_2$ and CH$_4$ ice \citep[see e.g. Fig. 3c in ][]{Eistrup16}.
Finally, volatile elements, such oxygen and carbon, may be locked up in large icy bodies in the midplane\citep{Bergin10,Ros13,Guidi16}. These large pebbles cannot diffuse upward and participate in the gas-phase chemistry \citep[see][]{Du15,Kama16}. Such a process is likely the cause of the under-abundance of gas-phase water in disks atmosphere.

\subsubsection{Test models}
\label{test_models}

From the modeling side, it is possible to simulate the two different scenarios (low gas-to-dust ratio or large carbon depletion) and to compare the predictions with the observations. As done by \cite{Miotello16}, the loss of gas is simulated by fixing the gas mass and increasing the dust mass, i.e. obtaining lower gas-to-dust ratios. The carbon depletion scenario is obtained reducing the initial ISM-like carbon abundance by different levels. We define carbon depletion as $\delta_{\rm C}$=1 if the carbon over hydrogen ratio is set to the ISM-level, [C]/[H] = $1.35 \cdot 10^{-4}$. We then assume higher values of carbon depletion $\delta_{\rm C}$=0.1, 0.01 if the abundance ratio is respectively [C]/[H] = $1.35 \cdot 10^{-5}$, $1.35 \cdot 10^{-6}$.

\begin{figure*}
   \resizebox{\hsize}{!}
             {\includegraphics[width=1.\textwidth]{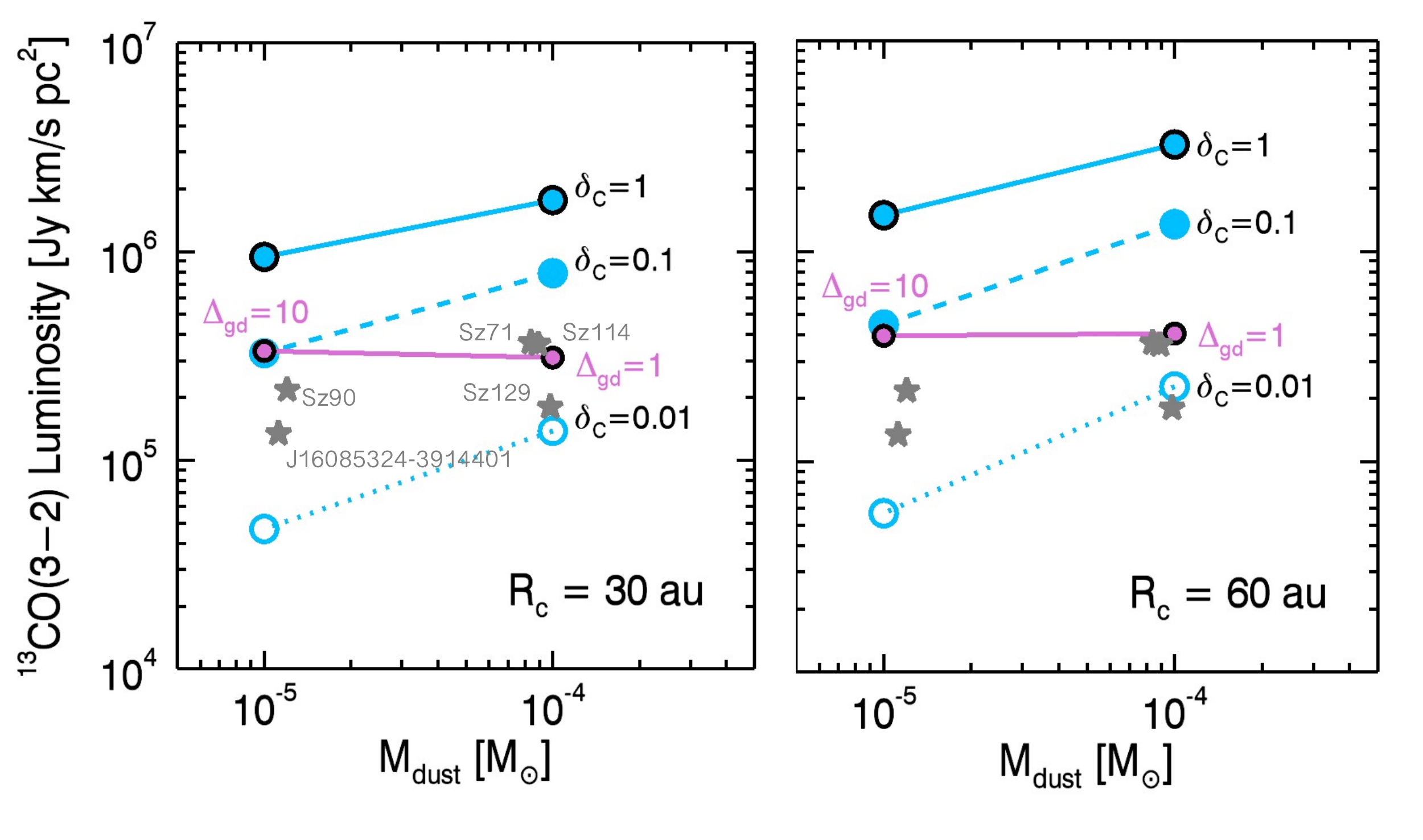}}
      \caption{$^{13}$CO ($J$=3-2) luminosity as a function of dust mass. Light blue circles present the simulated line luminosities with different levels of carbon depletion: no depletion ($\delta_{\rm C}=1$, black contoured circles), depletion by a factor of 10 ($\delta_{\rm C}=0.1$, filled light blue circles) and depletion by two orders of magnitude ($\delta_{\rm C}=0.01$, light blue empty circles). Purple circles present the simulated line luminosities obtained with no carbon depletion, fixing the gas mass $M_{\rm gas}=10^{-4} M_{\odot}$ and increasing the dust mass to $M_{\rm dust}=10^{-5}, 10^{-4}M_{\odot}$, accordingly with gas-to-dust ratios $\Delta_{\rm gd}$=10, 1. A subsample of six Lupus sources is shown by the gray stars. Models run with $R_{\rm c}$ = 30 au and with $R_{\rm c}$ = 60 au are shown in the left  and right panels respectively.}
       \label{C_depl_models}
\end{figure*}

Two interesting groups of sources can be identified in Fig. \ref{g_t_d} for which the gas mass is $M_{\rm gas}=10^{-4} M_{\odot}$, while the dust mass is either  $M_{\rm dust}=10^{-4}M_{\odot}$ (Sz71, Sz114, Sz129) or $M_{\rm dust}=10^{-5}M_{\odot}$ (J16085324-3914401, Sz90). These five disks present similar $^{13}$CO emission, but different continuum fluxes. This may be interpreted as the first group of sources presenting a lower gas-to-dust ratio, or a higher level of carbon depletion. Four additional models have been run with $R_{\rm c}=$ 30, 60 au, fixing the gas mass $M_{\rm gas}=10^{-4} M_{\odot}$ and increasing the dust mass from $M_{\rm dust}=10^{-6}M_{\odot}$ to $M_{\rm dust}=10^{-5}, 10^{-4}M_{\odot}$, i.e. with gas-to-dust ratios of 10 and 1 (see purple circles in Fig. \ref{C_depl_models}). Moreover, another eight models have been run with $R_{\rm c}=$ 30, 60 au, $M_{\rm dust}=10^{-5}, 10^{-4}M_{\odot}$, gas-to-dust ratios of 100, but with initial carbon abundances reduced by a factor  $\delta_{\rm C}$=0.1, 0.01 (see Fig. \ref{C_depl_models}). Remarkably, for a fixed gas mass and $\delta_{\rm C}$, the CO line intensities do not depend much on dust masses.

Comparing the results from the two sets of models, presented in Fig. \ref{C_depl_models}, with the observations it is not possible to rule out one of the two scenarios, pointing to the need to calibrate CO-based masses with other tracers. The first group of sources (Sz71, Sz114, Sz129), bright in the continuum but faint in CO emission, are well described either by models with gas-to-dust ratios equal to unity or with a carbon depletion of around two orders of magnitudes. The fluxes of the second set of disks (J16085324-3914401, Sz90) are instead reproduced with less precision by models with a gas-to-dust ratio equal to 10, or with carbon depletion around a factor of 10. Possibly, both carbon depletion and gas dissipation processes are playing a role, but it is not possible to estimate the relative importance. 

\subsection{Correlation between disk gas mass and stellar mass}
\label{Macc}
\begin{figure}
   \resizebox{\hsize}{!}
             {\includegraphics[width=1.\textwidth]{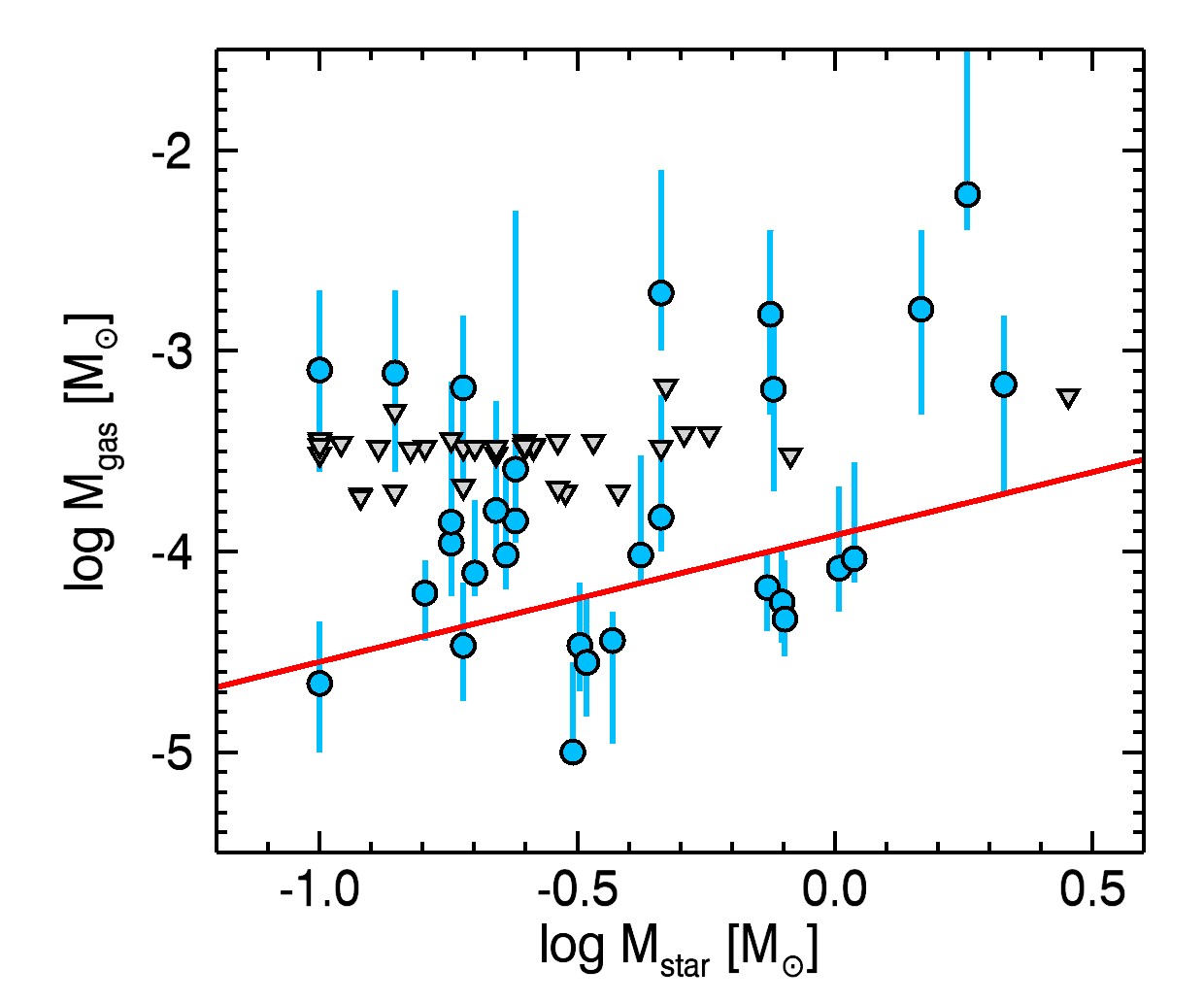}}
      \caption{Disk gas mass as a function of stellar mass for Lupus disks. Gas non-detections are shown as upper limits by gray triangles. For the gas detections the error bars are reported. The red line gives the Bayesian linear regression fit \citep{Kelly07}.}
       \label{Mstar}
\end{figure}

\cite{Ansdell16} found a positive correlation between $M_{\rm gas}$ and $M_{\star}$ \citep[computed by][2016 subm.]{Alcala14}, but were not able to make a meaningful fit given the low number of disks detected in both CO isotopologues and the large uncertainties on the gas mass determinations. This changes if the sample is enlarged from 10 to 34 sources. Using the Bayesian linear regression method \citep{Kelly07} which accounts for the upper limits, we find a correlation with $r = 0.74$ and a two-sided $p$-value of $3\cdot 10^{-2}$ for the null hypothesis that the slope of this correlation is zero. This is shown in  Fig. \ref{Mstar} where the red line gives the Bayesian linear regression fit, which considers errors on both axes and is applied to both detected and undetected targets.  The slope of the correlation is 0.63 and the intercept is -3.92.

\cite{Ansdell16} found a clear correlation between $M_{\rm dust}$ and $M_{\star}$. Furthermore, \cite{Manara16} found a linear relation between the mass accretion rate onto the central star and the disk mass inferred from the dust, combining ALMA and VLT/X-Shooter data \citep[][2016 subm.]{Alcala14}. A relation between $\dot{M}_{\rm acc}$ and $M_{\rm disk}$ has been theoretically predicted for viscously evolving disks \citep[][]{Hartmann98} and it was found observationally in Lupus for the first time using dust masses to measure the bulk disk mass, which is equivalent to assume a constant gas-to-dust ratio \citep{Manara16}. For a gas-to-dust ratio of 100, the ratio of $M_{\rm disk}/\dot{M}_{\rm acc}$ is comparable with the age of the region, as expected for viscous disks and under the effect of other disk evolution processes \citep{Jones12,Rosotti16}. On the other hand, no correlation was found with disk masses derived from CO isotopologues lines by \cite{Ansdell16}. Even expanding the sample of gas masses to 34 sources, no correlation is found in this work. Furthermore, the ratio of $M_{\rm disk}/\dot{M}_{\rm acc}$ when using $M_{\rm disk}$ derived from CO isotopologues is much smaller than the age of the region. This is expected only in cases where external disturbances, such as external photoevaporation, play a major role in the evolution of disks \citep{Rosotti16}. As this should not be the case in the Lupus region, this method suggests that CO-based  disk masses are generally under-estimated.

Assuming that all Lupus disks have evolved mainly due to viscous processes over the past few Myr, the observed correlation between the current mass accretion rate and dust mass found by \cite{Manara16} implies a constant gas-to-dust ratio, which is close to 100 based on the observed $M_{\rm disk}/M_{\rm acc}$ ratio. This in turn points to a scenario in which carbon depletion mechanisms are in play and affect CO line emission. 
Moreover, the non-correlation found between mass accretion rate and CO-based gas masses implies that carbon is non-homogeneously depleted throughout the Lupus sample. This opens an interesting question on the stellar and disk properties that maximize the sequestering of carbon from gas-phase CO. As discusses in Sect. \ref{C_depl}, if an anti-correlation of  complementary gas tracers, such \textbf{as} C$_2$H, with CO lines will be observed, this will help in solving the controversy.


\section{Summary and conclusion}

In this work we have employed the grid of physical-chemical models presented in \cite{Miotello16} to analyze continuum and CO isotopologues ALMA observations of Lupus disks \citep{Ansdell16}. Employing also the $^{13}$CO-only detections in the optically thin regime, disk gas masses have been calculated for a total of 34 sources, expanding the sample of 10 disks for which \cite{Ansdell16} could derive gas masses. As previously found, overall gas masses are very low, often smaller than $M_{\rm Jup}$. Accordingly, global gas-to-dust ratios are much lower than the expected ISM-value of 100, lying predominantly between 1 and 10.  Low CO-based gas masses and gas-to-dust ratios may indicate rapid loss of
gas, or alternatively carbon depletion. The latter may occur via sequestering of carbon from CO that is either locked into large icy bodies in the midplane, or converted in more complex molecules via either gas-phase chemistry or ice chemistry \citep{Aikawa96,Bergin14,Du15,Eistrup16,Kama16,Yu16}. However, current ALMA data alone cannot distinguish between these scenarios. We have simulated both of them, but model results do not allow us to rule out one of the two hypotheses, pointing to the need to calibrate CO-based masses with other tracers. Alternatively, exploitation of disk mass and mass accretion rate measurements simultaneously available in protoplanetary disks can give insight on the nature of low CO-based masses \citep{Manara16} and point to non-homogeneous carbon depletion throughout the Lupus sample. 
 
\begin{table}[tbh]
\begin{footnotesize}
\caption{Disk mass determination for all the detected sources in Lupus. Upper limits on the disk mass are also reported. Notation: $^{(-n)}$ indicates $10^{-n}$}
\label{tab_masses}
\centering
\begin{tabular}{rlll}
\toprule
Source name& $M_{\rm gas} [M_{\odot}]$ & $M_{\rm max} [M_{\odot}]$ &$M_{\rm min} [M_{\odot}]$ \\
\midrule
Sz65 & 6.435$^{(-4)}$ & 2.0$^{(-4)}$ &  1.5$^{(-3)}$\\
Sz66 & 1.000$^{(-5)}$ & 1.0$^{(-5)}$  & 2.8$^{(-5)}$\\
J15430131-3409153 & <3.560$^{(-4)}$ & - & - \\ 
J15430227-3444059 & <3.060$^{(-4)}$ & - & - \\
J15445789-3423392 & <1.872$^{(-4)}$ & - & - \\
J15450634-3417378 & 2.400$^{(-5)}$ & 1.3$^{(-5)}$ & 5.0$^{(-5)}$\\
J15450887-3417333 & 7.742$^{(-4)}$ & 2.5$^{(-4)}$  & 2.0$^{(-3)}$  \\
Sz68 & 6.800$^{(-4)}$ &  2.0$^{(-4)}$ &  1.5$^{(-3)}$\\
Sz69 & 3.400$^{(-5)}$ &  1.8$^{(-5)}$ &  7.0$^{(-5)}$ \\
Sz71 & 9.600$^{(-5)}$ &  7.0$^{(-5)}$ & 3.0$^{(-4)}$ \\
Sz72 & <1.965$^{(-4)}$ & - & - \\
Sz73 & <3.012$^{(-4)}$ & - & - \\
Sz74 & <2.055$^{(-4)}$ & - & - \\
Sz81 & <3.372$^{(-4)}$ & - & - \\
Sz83 & 1.5162$^{(-3)}$ &  4.8$^{(-4)}$ & 4.0$^{(-3)}$ \\
Sz84 & 1.100$^{(-4)}$ &  6.0$^{(-5)}$ & 2.2$^{(-4)}$ \\
Sz129 & 4.600$^{(-5)}$ &  3.0$^{(-5)}$ &  9.0$^{(-5)}$ \\
J15592523-4235066& <1.920$^{(-4)}$ & - & - \\
RYLup & 1.607$^{(-3)}$ &  4.8$^{(-4)}$ &  4.0$^{(-3)}$ \\ 
J16000060-4221567& <2.102$^{(-4)}$ & - & - \\
J16000236-4222145 & 1.420$^{(-4)}$ & 1.1$^{(-4)}$ & 7.0$^{(-4)}$ \\
J16002612-4153553& <1.965$^{(-4)}$ & - & - \\
Sz130 & 3.600$^{(-5)}$ &  1.1$^{(-5)}$ &  5.0$^{(-5)}$ \\
MYLup & 8.250$^{(-5)}$ & 5.0$^{(-5)}$ &  2.1$^{(-4)}$ \\
Sz131& <1.965$^{(-4)}$ & - & - \\
J16011549-4152351 & 2.374$^{(-3)}$ & 1.5$^{(-3)}$ &  5.0$^{(-2)}$ \\
Sz133 & 7.250$^{(-5)}$ &  3.0$^{(-5)}$ &  9.0$^{(-5)}$ \\
Sz88A& <3.860$^{(-4)}$ & - & - \\
Sz88B & <3.292$^{(-4)}$ & - & - \\
J16070384-3911113 & 8.022$^{(-4)}$ & 2.5$^{(-4)}$ & 2.0$^{(-3)}$ \\
J16070854-3914075 &2.025$^{(-4)}$ & 1.8$^{(-4)}$ & 4.0$^{(-3)}$ \\
Sz90 & 5.600$^{(-5)}$ &  3.500$^{(-5)}$ & 1.0$^{(-4)}$ \\
J16073773-3921388 & 3.455$^{(-4)}$ & - & - \\
Sz95 & 2.800$^{(-5)}$ &  1.5$^{(-5)}$ &  6.0$^{(-5)}$ \\
J16075475-3915446& <4.670$^{(-4)}$ & - & - \\
J16080017-3902595 & <3.292$^{(-4)}$ & - & - \\
J16080175-3912316 & <5.562$^{(-4)}$ & - & - \\
Sz96 & <3.292$^{(-4)}$ & - & - \\
J16081497-3857145 & 2.200$^{(-5)}$ &  1.0$^{(-5)}$ & 4.5$^{(-5)}$ \\
 Sz97 & <3.535$^{(-4)}$ & - & - \\
Sz98 & 6.600$^{(-5)}$ &  4.0$^{(-5)}$ & 1.0$^{(-4)}$ \\
 Sz99 & <3.050$^{(-4)}$ & - & - \\
Sz100 & 1.400$^{(-4)}$ &  1.1$^{(-4)}$ & 7.0$^{(-4)}$ \\
 J160828.1-391310 & <3.050$^{(-4)}$ & - & - \\
Sz103& <3.292$^{(-4)}$ & - & - \\
J16083070-3828268 & 6.01$^{(-3)}$ & 4.0$^{(-3)}$ &  1.0$^{(-1)}$ \\
Sz104 & <3.617$^{(-4)}$ & - & - \\
 J160831.1-385600 & <5.400$^{(-4)}$ & - & - \\
V856Sco & <5.970$^{(-4)}$ & - & - \\
Sz106 & <3.860$^{(-4)}$ & - & - \\
Sz108B  & 6.537$^{(-4)}$ &  2.0$^{(-4)}$ & 1.5$^{(-3)}$ \\
J16084940-3905393  & <3.292$^{(-4)}$ & - & - \\
V1192Sco & <3.292$^{(-4)}$ & - & - \\
Sz110 & <3.372$^{(-4)}$ & - & - \\
J16085324-3914401 & 3.400$^{(-5)}$ &  2.0$^{(-5)}$ &  7.0$^{(-5)}$ \\
J16085373-3914367  & <3.617$^{(-4)}$ & - & - \\
Sz111 & 1.935$^{(-3)}$ & 1.0$^{(-3)}$ & 8.0$^{(-3)}$ \\
J16085529-3848481 & <3.535$^{(-4)}$ & - & - \\
Sz112 & <3.292$^{(-4)}$ & - & - \\
Sz113 & <3.292$^{(-4)}$ & - & - \\
J16085828-3907355 & <5.077$^{(-4)}$ & - & - \\
J16085834-3907491 & <5.400$^{(-4)}$ & - & - \\
\midrule
\end{tabular}
\end{footnotesize}
\end{table}

\begin{table}[tbh]
\begin{footnotesize}
\centering
\begin{tabular}{rlll}
\toprule
J16090141-3925119 & 2.575$^{(-4)}$ &  1.5$^{(-4)}$ &  5.0$^{(-3)}$ \\
Sz114 & 9.600$^{(-5)}$ &  6.5$^{(-5)}$ &  2.8$^{(-4)}$ \\
Sz115 & <3.292$^{(-4)}$ & - & - \\
J16091644-3904438 & <5.562$^{(-4)}$ & - & - \\
J16092032-3904015 & <4.995$^{(-4)}$ & - & - \\
J16092317-3904074 & <5.400$^{(-4)}$ & - & - \\
J16092697-3836269 & 7.800$^{(-5)}$ &  6.0$^{(-5)}$ & 1.8$^{(-4)}$ \\
J160934.2-391513 & <5.562$^{(-4)}$ & - & - \\
J16093928-3904316 & <5.400$^{(-4)}$ & - & - \\
Sz117 & <3.535$^{(-4)}$ & - & - \\
Sz118 & 1.175$^{(-4)}$ &  7.0$^{(-5)}$ & 2.8$^{(-4)}$ \\
J16095628-3859518 & <3.372$^{(-4)}$ & - & - \\
J16100133-3906449 & <4.995$^{(-4)}$ & - & - \\
J16101307-3846165 & <3.455$^{(-4)}$ & - & - \\
J16101857-3836125 & <3.212$^{(-4)}$ & - & - \\
J16101984-3836065 & <3.372$^{(-4)}$ & - & - \\
J16102741-3902299 & <5.400$^{(-4)}$ & - & - \\
J16102955-3922144 & 1.600$^{(-4)}$ &  1.0$^{(-4)}$ & 5.6$^{(-4)}$ \\
J16104536-3854547 & <5.725$^{(-4)}$ & - & - \\
Sz123B & <3.535$^{(-4)}$ & - & - \\
Sz123A & 1.480$^{(-4)}$ & 1.0$^{(-4)}$ & 6.0$^{(-4)}$ \\
J16115979-3823383 & <3.212$^{(-4)}$ & - & - \\
J16120445-3809589 & <5.725$^{(-4)}$ & - & - \\
J16121120-3832197 & <5.645$^{(-4)}$ & - & - \\
J16124373-3815031 & <6.617$^{(-4)}$ & - & - \\
J16134410-3736462 & 6.200$^{(-5)}$ &  3.6$^{(-5)}$ &  9.0$^{(-5)}$ \\
\hline
\end{tabular}
\end{footnotesize}
\end{table}

\section*{Acknowledgements}

This  paper  makes use of the following ALMA data: ADS/JAO.ALMA\#2013.1.00220.S. ALMA is a partnership of ESO (representing its member states), NSF (USA) and  NINS (Japan), together with NRC (Canada), NSC and  SIAA (Taiwan), and KASI  (Republic  of  Korea), in cooperation with the Republic of Chile. The Joint ALMA Observatory is operated by ESO, AUI/
NRAO, and NAOJ. Astrochemistry in Leiden is supported by the Netherlands Research
School for Astronomy (NOVA), by a Royal Netherlands Academy of Arts
and Sciences (KNAW) professor prize, and by the European Union A-ERC
grant 291141 CHEMPLAN. 
This work was partly supported by the Italian Ministero dell’Istruzione, Universita e Ricerca through the grant Progetti Premiali 2012-iALMA (CUP C52I13000140001).
CFM gratefully acknowledges an ESA Research Fellowship. J.P.W. is supported by funding from the NSF and NASA through grants AST- 1208911 and NNX15AC92G.



\end{document}